\documentclass[12pt]{article} 

\newcommand{\beq}{\begin{eqnarray}}
\newcommand{\eeq}{\end{eqnarray}}
\newcommand{\rf}[1]{(\ref{#1})}

\begin{document}

\topmargin 0pt
\oddsidemargin 5mm
\headheight 0pt

\topskip 5mm

\thispagestyle{empty}
      
\begin{flushright}
NSF-KITP-05-19\\
BCCUNY-HEP/05-02\\
March 2005, Revised June 2005.
\hfill
\end{flushright}

\begin{center}

\hspace{10cm}

\vspace{15pt}
{\large \bf
ASYMPTOTIC FREEDOM OF ELASTIC STRINGS AND BARRIERS }

\vspace{20pt}

{\bf Peter Orland}$^{\rm a. b.c.}$\footnote{orland@gursey.baruch.cuny.edu, giantswing@gursey.baruch.cuny.edu}
and
{\bf Jing Xiao}$^{\rm b.c.}$\footnote{jxiao@gursey.baruch.cuny.edu}

\vspace{8pt}

\begin{flushleft}
a. Kavli Institute for Theoretical Physics, The University of California, Santa Barbara, CA 
93106, U.S.A.
 \end{flushleft}
 
\begin{flushleft}
b. Physics Program, The Graduate School and University Center,
The City University of New York, 365 Fifth Avenue,
New York, NY 10016, U.S.A.
\end{flushleft}

\begin{flushleft}
c. Department of Natural Sciences, Baruch College, The 
City University of New York, 17 Lexington Avenue, New 
York, NY 10010, U.S.A. 
\end{flushleft}

\vspace{40pt}

{\bf Abstract}
\end{center}

We study the problem of a quantized elastic string in the presence of an impenetrable wall. This
is a two-dimensional field theory of an $N$-component real scalar field $\phi$ which becomes interacting 
through
the restriction $\phi^{\rm T}
\phi \le \phi_{\rm max}^{2}$, for 
a spherical wall of radius $\phi_{\rm max}$. The $N=1$ case is a string vibrating in a plane between two straight walls. We 
review a simple nonperturbative
argument that there is a mass gap in the spectrum, with asymptotically-free behavior in the 
coupling $g=\phi_{\rm max}^{-1}$, for $N\ge 1$. This scaling behavior of the mass gap 
has been disputed in some of the recent literature. We find, however, that
perturbation theory and the $1/N$ expansion each 
confirms that these models are asymptotically free. The $N\rightarrow \infty$ limit 
coincides with that of the O($N$) nonlinear sigma model. A 
$\theta$ parameter and instantons exist for the two-dimensional $N=2$ 
model, which describes a string confined to the interior of a cylinder
of radius $\phi_{\rm max}$.

\vfill
\newpage
\pagestyle{plain}
\setcounter{page}{1}

\section{Introduction}
\setcounter{equation}{0}
\renewcommand{\theequation}{1.\arabic{equation}}

We will make a few observations in this paper concerning a nonrelativistic elastic
string in $N$ transverse dimensions 
in the presence of barriers. This is a theory of an $N$-component scalar field
\begin{eqnarray}
\phi = \left( \begin{array}{cc} \phi_{1}\\
\phi_{2}\\
\cdot \\
\cdot \\
\cdot \\
\phi_{N} \end{array} \right)\;,
\nonumber
\end{eqnarray}
satisfying the condition  $\phi^{\rm T}\phi\le \phi_{\rm max}^{2}$.  The quantum string has the 
Lagrangian 
${\cal L}=
\frac{1}{2}\partial_{t}\phi^{\rm T}\partial_{t}\phi
-\frac{1}{2}\partial_{x}\phi^{\rm T}\partial_{x}\phi$, where $\partial_{t}=\partial/\partial t$,
$\partial_{x}=\partial/\partial x$ and the superscript T denotes the transpose. This 
is not a free two-dimensional field theory, 
however, due
to the constraint. Here $\phi_{\rm max}$ is the radius
of an infinitely-deep spherical well and we interpret its reciprocal $g=\phi_{\rm max}^{-1}$, as the coupling
constant. We shall refer to these as the $B^{N}$ models, since target space is the $N$-dimensional
ball
$B^{N}$. 

Note the similarity of the $B^{N}$ model to the 
$O(N)$ nonlinear sigma 
model, for
which the Lagrangian is the same, but the constraint is 
$\phi^{\rm T}\phi=1/g^{2}$. Indeed for large dimension
of target space $N$, most points of an $N$-dimensional ball are concentrated near the 
boundary. For this
reason, one expects the two models to coincide as 
$N\rightarrow \infty$.  In the large-$N$ limit, it makes little difference whether the constraint is
$\phi^{T}\phi=g^{-2}$ or
$\phi^{T}\phi\le g^{-2}$.

The $N=1$ model is of particular interest. This is a quantum string 
constrained to move in a planar channel of
width $2/g$. The model arises in the statistical mechanics of a two-dimensional membrane
between planar walls separated by a distance $2/g$ \cite{janke-kleinert}, \cite{sorn}, 
as 
well as striped phases of copper-oxide layers \cite{zaanen}, \cite{kivelson} .  Another motivation
for studying fields restricted in this way has been given in reference \cite{li-meurice}, where it has been argued 
that it has practical utility in perturbation expansions.

A simple argument repeated below shows that the spectrum
has a mass gap $M$, with the behavior
\begin{eqnarray}
M\simeq
\exp-Ag^{-2},\;g\rightarrow 0\;,
\label{asymp-free}
\end{eqnarray} 
where $A$ is a constant. This result was known to, but  disputed by 
Zaanen et. al. \cite{zaanen1},  who
use an argument similar to that of Helfrich and Servuss \cite{helfrich1} to conclude
that
\begin{eqnarray}
M\simeq \exp-Ag^{-\alpha} ,\; g \rightarrow 0\;,
\label{not-asymp-free}
\end{eqnarray}
where $\alpha\approx 2/3$. Nishiyama 
studied the model with the density-matrix renormalization group and
has also argued for (\ref{not-asymp-free}) \cite{nishiyama}. 

We show in this paper is that \rf{asymp-free} is correct according to
standard analytic methods. We consider two such 
methods. The first of these is a
simple one-loop renormalization group 
analysis for $N=1$. The other is the $1/N$ expansion. In any case, if
one accepts (\ref{not-asymp-free}), the inevitable conclusion is that the analytic part of the
beta function is zero - which we show is not the case.

We also consider the interesting case of $N=2$. This field theory describes a string allowed to
vibrate inside a cylinder. We find that there is a topological term which can be included in
the action and that there are instantons. We determine the instanton solutions; they are similar to those of the circular brane
model \cite{lukyanov}.

Before concluding this introduction, we repeat the argument given in references 
\cite{janke-kleinert},
\cite{sorn},
\cite{zaanen1}, and \cite{nishiyama}  that a mass gap appears
and depends on the coupling as (\ref{asymp-free}). Though referred to as a mean-field argument in
\cite{nishiyama}, it is closer in spirit to the theorems of Peierls and of Mermin and Wagner forbidding
continuous symmetry breaking in two dimensions. If we ignore the constraint and use our massless 
Lagrangian, the two-point equal-times correlation function behaves as
\begin{eqnarray}
\left<\phi^{T}(x) \phi(0)
\right>=
\frac{N}{2\pi} \ln \frac{\vert x \vert}{a}\;, \label{massless}
\end{eqnarray}
where $a$ is a short-distance cut-off. But we also have the strict inequality
\begin{eqnarray}
\left<\phi^{T}(x) \phi(0)
\right> \le g^{-2}\;. \nonumber
\end{eqnarray}
Thus, (\ref{massless}) should be valid for $\vert x \vert$ approximately in the 
range $a<\vert x\vert< a e^{\frac{2\pi}{g^{2}N} }$, but not for 
$\vert x\vert > a e^{\frac{2\pi}{g^{2}N}} $. This limiting
value of $\vert x \vert$ should be the correlation length of the theory, which is the inverse of the mass
gap. Thus we obtain 
\begin{eqnarray}
M\simeq
a^{-1}\exp-\frac{2\pi}{g^{2}N},\;g\rightarrow 0 \;,
\label{correct-asymp-free}
\end{eqnarray} 
which is just (\ref{asymp-free}) with $A=2\pi/N$. We shall verify (\ref{correct-asymp-free})
in the next section. This argument is very 
suggestive - perhaps it will point the way to a
rigorous proof of the mass gap, which is lacking for many interesting field 
theories. The reader should take
the argument with a grain of salt, however. It 
does not really establish that the two-point function falls off 
exponentially, but only shows
that it cannot behave logarithmically. For example, power-law 
decay of the two-point function cannot easily be ruled
out.

The beta function to lowest order follows simply from (\ref{correct-asymp-free}). It is negative and
vanishes at $g=0$:
\beq
\beta(g)=\left. \frac{\partial g^{2} }{\partial \ln a^{-1}}\right\vert_{M \;{\rm fixed}}=-\frac{g^{4}N}{2\pi}\;   . \nonumber
\eeq

\section{Expansion methods}
\setcounter{equation}{0}
\renewcommand{\theequation}{2.\arabic{equation}}

The Lagrangian of the $N=1$ model, Wick rotated to Euclidean space is
\begin{eqnarray}
{\cal L}=
\frac{1}{2}\partial_{\mu}\phi\partial^{\mu}\phi   \;,   \nonumber
\end{eqnarray}
where $\partial_{\mu}=\partial/\partial x^{\mu}$, where 
$x^{0}=t$, $x^{1}=x$. Since $\phi^{2}\le g^{-2}$, we parametrize $\phi$ by a new field $\psi$, through
$\phi(x)=g^{-1}\sin\psi(x)$. This choice of  parametrization is not unique (other choices of
parametrization,, such as 
$\phi=g^{-1}\tanh \psi$ would give the same result). This mapping from of $\psi$ to $\phi$ is
many-to-one, instead of one-to-one, but this fact will not make any difference as far
as perturbation theory is concerned.

The Lagrangian becomes
\begin{eqnarray}
{\cal L} & = & \frac{1}{2g^{2}}\cos^2 \psi \; \partial_{\mu}\psi\partial^{\mu}\psi  \nonumber \\
         & = & \frac{1}{2g^{2}}(1-\psi^2+\frac{1}{3} \psi^4)\partial_{\mu}\psi\partial^{\mu}\psi\;. 
\nonumber
%         & = & \frac{1}{2}(1-{\psi'^2}g^{2}+\frac{1}{3} {\psi'^4)}g^{4}\partial_{\mu}\psi'\partial^{\mu}\psi' \;,  %\nonumber
\end{eqnarray}
%Where we expand $\cos\psi$, and rescale $\psi$ to $\psi'=g{-1}\psi$.
The functional integral is 
\begin{eqnarray}
W[J]=\int [d\psi] \exp{-\frac{1}{\hbar}\int d^{2} x[{\cal L}-J\psi-\hbar a^{-2}\ln (\cos\psi)]} \;, \nonumber
\end{eqnarray}
where $a$ is a short-distance cut-off with dimensions of centimeters. The third term in the exponent,
which comes from the Jacobian in the functional measure, is of one higher order of $\hbar$ order than the Lagrangian; though this term must be considered at two loops, we may ignore it
in our one-loop calculation.
 
%The 1-loop contribution to the effective action is given by the diagrams
%\begin{eqnarray}
%{Feynman diagram1?}&=&\frac{2}{2d^2}\int\frac{dq^2}{(2\pi)^2}\frac{1}{q^2}\\
%                   &=&\frac{1}{2\pi d^2}\ln\frac{\Lambda ^2}{\mu ^2}  \;, \nonumber
%\end{eqnarray}
%and
%\begin{eqnarray}
%{Feynman diagram2?}&=&\frac{2}{2d^2}\int\frac{dq^2}{(2\pi)^2}\frac{q^2}{q^2}\\
%				   &=&\frac{1}{4\pi d^2}(\Lambda ^2-\mu ^2)\;, \nonumber
%\end{eqnarray}
%
The leading term of the 
effective action is
\begin{eqnarray}
\frac{1}{2g^{2}}\left[1-\frac{g^{2}}{4\pi}\ln({\mu ^2}a^{2})\right]\partial_{\mu}\phi\partial^{\mu}\phi\;,
\nonumber
\end{eqnarray}
where $\mu$ is an infrared cut-off, with dimensions of inverse centimeters, and where a quadratically-divergent contribution is canceled by a counterterm. From this expression
we find that the mass gap scales as
\begin{eqnarray}
M\simeq
a^{-1}\exp-\frac{2\pi}{ g^{2}}\;. \nonumber
\end{eqnarray}
This agrees with the result (\ref{correct-asymp-free}) discussed
in the introduction.

The $1/N$ expansion for the $B^{N}$ model is extremely simple. At leading order, all expressions
coincide with those of the nonlinear O($N$) sigma model. There is some difference to first order
in $1/N$, but we will not discuss this issue in detail.

Using a standard integral formula for the Heaviside function, the functional integral
is
\beq
Z=\int  [d\omega] [d\phi] \exp-\int  d^{2}x \left[\frac{1}{2}\partial_{\mu}\phi^{\rm T}\partial_{\mu}\phi
+{\rm i}\omega(\phi^{\rm T} \phi-G^{-2}N)+a^{-d}\ln (\omega-{\rm i}\epsilon)\right]\;,
\nonumber
\eeq
where $G^{2}=g^{2}/N$. Integration over $\phi$ yields
\beq
Z=\int [d\omega]\exp -N\left[ \frac{1}{2}{\rm Tr} \ln(-\partial^{2}/2-{\rm i}\omega)
-{\rm i} G^{-2} \int d^{2}x \; \omega +\frac{1}{N} a^{-d}\ln(\omega-{\rm i}\epsilon)\right] \;.
\label{determinant}
\eeq
This integral is dominated by a saddle point on the imaginary axis for large $N$, which
we write as $\omega_{0}=-{\rm i}\frac{m^{2}}{2}$. The presence of $\epsilon$ in (\ref{determinant}) assures that the logarithm is defined on the correct sheet in the
vicinity of the saddle point.  Except for the last term in (\ref{determinant}) this is the same expression
obtained for the sigma model. The equation for the saddle point is
\beq
1=G^{2}\int\frac{d^{d}p}{(2\pi)^{d}}\frac{1}{p^{2}+m^{2}} \;. \label{gap}
\eeq
If the momentum integral (\ref{gap}) is cut off by $\vert p \vert<a^{-1}$, we find the standard result
\beq
m=a^{-1}e^{-\frac{2\pi}{G^{2}}}\left[ 1-e^{-\frac{4\pi}{G^{2}}}\right]^{-1/2}\;,
\nonumber
\eeq
confirming that 
the model is asymptotically free.

The $B^{N}$ model is different from the $O(N)$ sigma model to order $1/N$. That is because the last
term in (\ref{determinant}) will contribute to this order.

\section{The topological term and instantons for $N=2$}
\setcounter{equation}{0}
\renewcommand{\theequation}{3.\arabic{equation}}

For the case of $N=2$ in two dimensions, a new term can be added to the 
action. We consider the $B^{2}$ model
in Euclidean space-time. We take this 
space-time to be a 
two-dimensional ball, i.e. a disk, with radius $R$. The target space is also a disk 
with radius $g^{-1}$. A smooth field
configuration is a map from the first disk to the second. This map has a degree 
which is the number of
images of the space-time disk in the target space disk. This degree is
\begin{eqnarray}
\nu=\frac{g^{2}}{2\pi} \int_{\vert x\vert\le R} d^{2}x 
\;(\partial_{1}\phi^{1} \partial_{2}\phi^{2}
-\partial_{1}\phi^{2} \partial_{2}\phi^{1})\;. \nonumber
%\label{degree}
\end{eqnarray}
The action with such a term is
\begin{eqnarray}
S=S_{0}+{\rm i}\theta \nu=\frac{1}{2} \int_{\vert x\vert\le R}  d^{2}x\;(
\partial_{1}\phi^{T}\partial_{1}\phi+
\partial_{2}\phi^{T}\partial_{2}\phi)\;+\;{\rm i} \theta \nu\;, \nonumber
%\label{topological-action}
\end{eqnarray}
where the (time-reversal violating) parameter $\theta$ is defined modulo $2 \pi$.

A similar term can be defined for the circular-brane model \cite{lukyanov}. In this model, the 
field $\phi$ is unconstrained except at the
boundary, where it is required that $\phi^{T}\phi=g^{-2}$. The instanton solutions are of the
same form as those found below.

%It is straightforward to prove an inequality between the first term $S_{0}$ in the action and
%the degree of the mapping, as has been done for other models. We have two algebraic inequalities
%\begin{eqnarray}
%(\partial_{1}\phi^{1}\mp\partial_{2}\phi^{2})^{2}
%+(\partial_{2}\phi^{1}\pm\partial_{1}\phi^{2})^{2}\ge 0\;. \nonumber
%\end{eqnarray}
%These immediately yield
%\begin{eqnarray}
%S_{0}\mp 2\pi g^{-2}\nu\ge 0\;, \nonumber
%\end{eqnarray}
%hence
%\begin{eqnarray}
%S_{0}\ge 2\pi g^{-2}\vert \nu \vert\;. \label{topological-inequality}
%\end{eqnarray}

%The inequality (\ref{topological-inequality}) is saturated for instantons 
%\begin{eqnarray}
%\partial_{1}\phi^{1}=\partial_{2}\phi^{2}\;,\;\;
%\partial_{2}\phi^{1}=-\partial_{2}\phi^{1} \;, \label{instantons}
%\end{eqnarray}
%with $\nu$ positive, as well as anti-instantons
%\begin{eqnarray}
%\partial_{1}\phi^{1}=-\partial_{2}\phi^{2}\;,\;\;
%\partial_{2}\phi^{1}=\partial_{2}\phi^{1} \;, \label{anti-instantons}
%\end{eqnarray}
%with $\nu$ negative. Notice that 
%the equation of motion 
%$(\partial_{1}^{2}+\partial_{2}^{2})\phi=0$ results from the imposition of either (\ref{instantons}) or 
%(\ref{anti-instantons}). Though the equation 
%of motion is linear, the nontrivial constaint $\phi^{\rm T}\phi\le g^{-2}$ is also obeyed, 
%as we shall show
%in a moment.

To show explicitly all the instantons and anti-instantons, we use complex coordinates. Let us define
$z=x^{1}+{\rm i}x^{2}$, ${\bar z}=x^{1}-{\rm i}x^{2}$, 
$\partial=\frac{1}{2}\partial_{1}-\frac{{\rm i}}{2}\partial_{2}=\partial/\partial z$
${\bar \partial}=\frac{1}{2}\partial_{1}+\frac{{\rm i}}{2}\partial_{2}=\partial / \partial{ \bar z}$
and 
$\phi=\phi^{1}+{\rm i}\phi^{2}$, ${\bar \phi}=\phi^{1}-{\rm i}\phi^{2}$. 
%Then 
%(\ref{instantons}) 
%implies that 
For instantons, $\phi$ is analytic and $\bar \phi$ is anti-analytic 
\begin{eqnarray}
\partial{\bar \phi}=0\;,\;\;{\bar \partial} \phi=0 \;, \nonumber
%\label{instantons1}
\end{eqnarray}
and
%(\ref{anti-instantons})
%implies that 
for anti-instantons, $\phi$ is anti-analytic and $\bar \phi$ is analytic 
\begin{eqnarray}
\partial\phi=0\;,\;\;{\bar \partial} {\bar \phi}=0\;.  \nonumber
%\label{anti-instantons1}
\end{eqnarray}

The general instanton solution must be an analytic map from the disk of radius $R$ to the disk
of radius $g^{-1}$ of degree $\nu$. This can only be of the form
\begin{eqnarray}
\phi=g^{-1}\prod_{j=1}^{\nu}\frac{a_{j}z+b_{j}R}{{\bar b_{j}}z+{\bar a_{j}}R} \;,\label{instanton-soln.}
\end{eqnarray}
with ${\bar \phi}$ given by complex conjugation. The complex moduli $a_{1}, \dots, a_{\nu}$ and $b_{1},\dots,b_{\nu}$
satisfy $\vert a_{j}\vert^{2}-\vert b_{j}\vert^{2}=1$. There are no poles 
in $z$ in the disk. To see this, note that the poles of (\ref{instanton-soln.}) lie at $z=- Ra_{j}/b_{j}$, $j=1,\dots, \nu$. Since 
$\vert a_{j}/b_{j}\vert^{2}=1+1/\vert b_{j} \vert^{2} \ge 1$, the function $\phi$ is completely analytic in the
interior of the disk. Furthemore, there is no singularity at the boundary at which $\vert \phi\vert =g^{-1}$. By 
the maximum-modulus theorem, our constraint is satisfied; $\vert \phi \vert $ 
cannot
exceed $g^{-1}$ anywhere in the disk of radius $R$.

The anti-instanton solutions are very similar to (\ref{instanton-soln.}). They are 
\begin{eqnarray}
\phi=g^{-1}\prod_{j=1}^{\nu}\frac{a_{j}{\bar z}+b_{j}R}{{\bar b_{j}}{\bar z}+{\bar a_{j}}R} \;
,\label{anti-instanton-soln.}
\end{eqnarray}
with ${\bar \phi}$ again given by complex conjugation. As before, the complex moduli $a_{1}, \dots, a_{\nu}$ and 
$b_{1},\dots,b_{\nu}$
satisfy $\vert a_{j}\vert^{2}-\vert b_{j}\vert^{2}=1$.

The semiclassical expansion about instantons is insufficient to understand the exponential
decay of
correlation functions. The fluctuation determinant is completely insensitive to both $\nu$ and the
moduli. An interesting question is whether fractional-topological charges can account for 
the correct exponentially-decaying behavior of correlation functions.

\section{Conclusion}

We have shown by several elementary methods that the energy gap in a quantum elastic
string with barriers decays exponentially with the square of the barrier width.  For the special
case of a string in the interior of a cylindrical barrier, there are instantons, which are holomorphic
maps from the two-dimensional disk to itself.

The
large-$N$ limit coincides with the spherical model in any dimension. The latter model has an 
ultraviolet-stable fixed point in three Euclidean dimensions (or two space and one time
dimensions). We expect that such a fixed point 
exists for the three-dimensional $B^{N}$ model for finite $N$, separating a spontaneously broken phase from
a massive, strong-coupling phase. In this case, the model describes a quantum
two-dimensional membrane, moving transversely,  in the presence of a barrier.

A new 
feature of the two-dimensional $B^{N}$ model is that the beta function is proportional to $N$; in
the case of the sigma model, the beta function is proportional to $N-2$ \cite{polyakov}.

\section*{Acknowledgments}

We thank Y. Meurice for discussions and for a critical reading of the manuscript. This research was
supported in part by the National Science Foundation under grant  No. PHY99-07949. It was also
supported in part by a grant from the PSC-CUNY.

\end{document}